\documentclass{appolb}
\usepackage{epsfig}

\usepackage{bm}%
\usepackage[colorlinks=true,linkcolor=red]{hyperref}
\usepackage{amsmath}
\usepackage{amsfonts}
\usepackage{amssymb}
\usepackage{cancel}
\usepackage{eucal}
\usepackage{hyperref}

\newcommand{\newc}{\newcommand}
\newc{\beq}{\begin{equation}}
\newc{\eeq}{\end{equation}}
\newc{\kt}{\rangle}
\newc{\bra}{\langle}
\newc{\beqa}{\begin{eqnarray}}
\newc{\eeqa}{\end{eqnarray}}
\newc{\pr}{\prime}
\newc{\longra}{\longrightarrow}
\newc{\ot}{\otimes}
\newc{\rarrow}{\rightarrow}
\newc{\h}{\hat}
\newc{\bom}{\boldmath}
\newc{\btd}{\bigtriangledown}
\newc{\al}{\alpha}
\newc{\be}{\beta}
\newc{\ld}{\lambda}
\newc{\ldmin}{\lambda_{\rm min}}
\newc{\sg}{\sigma}
\newc{\p}{\psi}
\newc{\eps}{\epsilon}
\newc{\om}{\omega}
\newc{\mb}{\mbox}
\newc{\tm}{\times}
\newc{\hu}{\hat{u}}
\newc{\hv}{\hat{v}}
\newc{\hk}{\hat{K}}
\newc{\ra}{\rightarrow}
\newc{\non}{\nonumber}
\newc{\ul}{\underline}
\newc{\hs}{\hspace}
\newc{\longla}{\longleftarrow}
\newc{\ts}{\textstyle}
\newc{\f}{\frac}
\newc{\df}{\dfrac}
\newc{\ovl}{\overline}
\newc{\bc}{\begin{center}}
\newc{\ec}{\end{center}}
\newc{\dg}{\dagger}
\newc{\prh}{\mbox{PR}_H}
\newc{\prq}{\mbox{PR}_q}

\begin{document}
\title{Largest Schmidt eigenvalue of entangled random pure states and conductance distribution in chaotic cavities.%
}
\author{Pierpaolo Vivo
\address{Abdus Salam International Centre for
Theoretical Physics\\
 Strada Costiera 11,
 34151 Trieste,
 Italy} 
} 

\maketitle
\begin{abstract}
A strategy to evaluate the distribution of the largest Schmidt eigenvalue for entangled random pure states 
of bipartite systems is proposed. We point out that the multiple integral defining the sought quantity for a bipartition of sizes $N,M$ is formally
identical (upon simple algebraic manipulations) to the one providing the probability density of Landauer conductance in open chaotic
cavities supporting $N$ and $M$ electronic channels in the two leads. Known results about the latter can then be straightforwardly employed in the former problem
for both systems with broken ($\beta=2$) and preserved ($\beta=1$) time reversal symmetry. The analytical results, yielding a continuous but not everywhere analytic distribution, are in excellent agreement with numerical simulations.
\end{abstract}
\PACS{03.67.Mn, 02.10.Yn, 02.50.Sk, 05.45.Mt, 73.23.-b, 21.10.Ft, 24.60.-k 
}

\section{Introduction}
Consider two sets of $n$ correlated random variables in $[0,1]$, $\{\lambda_i\}$ and $\{T_i\}$ ($i=1,\ldots,n$), respectively distributed according to the following joint probability densities (jpd):
\begin{align}
\mathcal{P}_1(\lambda_1,\ldots,\lambda_n) &=C_{n,\alpha}^{(\beta)}\delta\left(\sum_{i=1}^n\lambda_i -1\right)\prod_{i=1}^n
\lambda_i^{\alpha}\prod_{j<k}|\lambda_j-\lambda_k|^\beta\label{P1}\\
  \mathcal{P}_2(T_1,\ldots, T_n) &=K_{n,\alpha^\prime}^{(\beta)}\prod_{i=1}^n
  T_i^{\alpha^\prime}\prod_{j<k}|T_j-T_k|^\beta\label{P2}
\end{align}
where $C_{n,\alpha}^{(\beta)}$ and $K_{n,\alpha^\prime}^{(\beta)}$ are known normalization constants and $\beta=1,2$. In the following two subsection we will provide physical motivations for considering such sets, namely set $1$ corresponds to the distribution of Schmidt eigenvalues for
entangled random pure states in bipartite systems (see subsection \ref{ent}), while set $2$ corresponds to the distribution of transmission eigenvalues of an open cavity in the chaotic regime (see subsection \ref{condcav}).

Consider now the following statistical quantities:
\begin{itemize}
\item The cumulative distribution $Q_n(x)=\mathrm{Prob}[\lambda_{\mathrm{max}}\leq x]$ of the largest member of set $1$, $\lambda_{\mathrm{max}}=\max_i\{\lambda_i\}$. By definition, this is given by the following $n$-fold integral:
\begin{equation}\label{cum1}
Q_n(x)=\int_{[0,x]^n}d\lambda_1\cdots d\lambda_n \mathcal{P}_1(\lambda_1,\ldots,\lambda_n)
\end{equation}
Differentiating $Q_n(x)$, one obtains the \emph{probability density} of $\lambda_{\mathrm{max}}$, 
\beq
p_n(x)=\frac{d}{dx}Q_n(x)\label{densq}
\eeq
\item The probability density $\mathcal{P}_G(y)=\mathrm{Prob}[y\leq G\leq y+dy]$ of the quantity $G=\sum_{i=1}^n T_i$, which is given by the following $n$-fold integral
\footnote{One has the following normalizations $\int_{1/n}^1 dx\ p_n(x)=\int_0^n dy\ \mathcal{P}_G(y)=1$.}:
\begin{equation}
\mathcal{P}_G(y)=\int_{[0,1]^n}dT_1\cdots dT_n \mathcal{P}_2(T_1,\ldots, T_n)\delta\left(y-\sum_{i=1}^n T_i\right)
\end{equation}
\end{itemize}
 Simple algebraic manipulations, summarized in Appendix \ref{appA} lead to the following relation between the two quantities above:
 \begin{equation}\label{main}
 \boxed{Q_n(x)=\frac{C_{n,\alpha}^{(\beta)}}{K_{n,\alpha}^{(\beta)}}
 x^{n+\alpha n+\frac{\beta}{2}n(n-1)-1}\mathcal{P}_G\left(\frac{1}{x}\right),\qquad 1/n\leq x\leq 1,\ \alpha=\alpha^\prime}
 \end{equation}
The identity in eq. \eqref{main} is the main result of this paper\footnote{The bound $x\geq 1/n$ will be discussed in detail later on.}. Notwithstanding its remarkable simplicity, eq. \eqref{main} actually permits an exact evaluation of $Q_n(x)$, the so far unavailable distribution of the largest Schmidt eigenvalue for {\em finite} $n$ (see subsection \ref{ent}), in terms of $\mathcal{P}_G(y)$ (the probability density of Landauer conductance, see subsection \ref{condcav}) about which much more is known. 

The plan of the paper is as follows. In the next two subsections, we give a rather detailed introduction to the physics of entangled random pure states in bipartite systems (related to the set $1$ above) and about Landauer conductance in chaotic mesoscopic cavities supporting a finite number of electronic channels in the two attached leads (related to the set $2$ above). In section \ref{mainresults} we exploit the identity \eqref{main} to derive analytically the cumulative distribution and the density of $\lambda_{\mathrm{max}}$ in a few illustrative cases. These results are then compared with numerical simulations with excellent agreement. Eventually we present concluding remarks in section \ref{concl}
and technical developments in the three appendices.

\subsection{Entangled random pure states.}\label{ent}
Entanglement of pure bipartite systems is one of the most active areas of research nowadays, due to possible applications to quantum information and quantum computation problems
\cite{NeilsenBook,PeresBook}. It is also probably the simplest setting where well-behaved entanglement quantifiers can be defined, such as the von-Neumann or R\'enyi entropies of either subsystem \cite{PeresBook}, the so
called concurrence for two-qubit systems~\cite{Wootters} or other entanglement monotones \cite{cappellini,gour}. 

\emph{Typical} properties
of such states are best addressed by considering \emph{random} pure states (see e.g. \cite{majreview} for an excellent review).
More precisely, consider a bipartition of a $NM$-dimensional Hilbert space ${\cal 
H}^{(NM)}$ as ${\cal H}^{(NM)}={\cal H}^{(N)}_A \otimes {\cal H}^{(M)}_B$, where we assume
without loss of 
generality that $N\le M$.
For example, $A$ may be taken as a given
system (say a set of spins) living in an external environment (e.g., a 
heat bath) $B$.
A quantum state $|\psi\kt$ of the composite system can be expanded as a
linear combination
\begin{equation}\label{psi111}
|\psi\kt= \sum_{i=1}^N\sum_{\alpha=1}^M x_{i,\alpha}\, 
|i^A\kt\otimes |\alpha^B\kt
\end{equation}
where $|i^A\kt$ and $|\alpha^B\kt$ are two
complete basis of ${\cal H}^{(N)}_A $ and ${\cal H}^{(M)}_B$ respectively.
The coefficients $x_{i,\alpha}$'s of this expansion form the entries of a rectangular $(N\times M)$ matrix $\mathcal{X}$.

We consider here \emph{entangled random pure} states $|\psi\kt$. This means that: 
\begin{enumerate}
\item $|\psi\kt$ \emph{cannot} be expressed as a direct 
product
of two states belonging to the two subsystems $A$ and $B$. 
\item the expansion coefficients
$x_{i,\alpha}$ are random variables drawn from a certain probability distribution.   
\item the density matrix of the composite system is simply given by $\rho=|\psi\kt \bra 
\psi|$ with the constraint ${\rm Tr}[\rho]=1$, or equivalently $\bra \psi|\psi\kt=1$. 
\end{enumerate}

More precisely, the density matrix of $|\psi\kt$ can then be straightforwardly expressed as
\beq
\rho = \sum_{i,\alpha}\sum_{j,\beta} x_{i,\alpha}\, x_{j,\beta}^*\, |i^A\kt\bra j^A|\otimes 
|\alpha^B\kt 
\bra\beta^B|,
\label{dem2}
\eeq
where the Roman indices $i$ and $j$ run from $1$ to $N$ and the Greek indices $\alpha$ and 
$\beta$ run from $1$ to $M$.  

The \emph{reduced} density matrix $\rho_A= {\rm Tr}_B[\rho]$ is obtained by tracing out the environmental degrees of freedom (i.e. those of the subsystem $B$):
\beq
\rho_A = {\rm Tr}_B[\rho]=\sum_{\alpha=1}^M \bra \alpha^B|\rho|\alpha^B\kt.
\label{rdm1}
\eeq
Using the expansion in Eq. \eqref{dem2} one gets
\beq
\rho_A = \sum_{i,j=1}^N \sum_{\alpha=1}^M x_{i,\alpha}\, x_{j,\alpha}^*\, |i^A\kt\bra
j^A|=\sum_{i,j=1}^N W_{ij} |i^A\kt\bra j^A|
\label{rdm2}
\eeq
where $W_{ij}$'s are the entries of the $N\times N$ matrix $\mathcal{W}=\mathcal{X} \mathcal{X}^{\dagger}$.
In a similar way, one could obtain the reduced density matrix $\rho_B={\rm Tr}_A[\rho]$ of the 
subsystem $B$ in terms of the $M\times M$ matrix $\mathcal{W}^\prime=\mathcal{X}^\dagger \mathcal{X}$. The two matrices $\mathcal{W}$ and $\mathcal{W}^\prime$ share
the same set of nonzero (positive) real eigenvalues $\{\lambda_1,\lambda_2,\ldots,\lambda_N\}$.
In the diagonal basis, one can express $\rho_A$ as
\beq
\rho_A= \sum_{i=1}^N \lambda_i \, |\ld^A_i\kt\, \bra \ld^A_i|
\label{diagA}
\eeq
where $|\ld^A_i \kt$'s are the normalized eigenvectors of $\mathcal{W}=\mathcal{X}\mathcal{X}^{\dagger}$ and similarly for $\rho_B$.
The original composite state $|\psi\kt$ in this diagonal basis reads:
\beq
|\psi\kt = \sum_{i=1}^{N} \sqrt{\ld_i}\, |\ld_i^A\kt \otimes |\ld^B_i \kt
\label{Sch1}
\eeq
Eq. \eqref{Sch1} is known as the Schmidt decomposition, and the 
normalization
condition $\bra \psi|\psi\kt=1$, or equivalently ${\rm Tr}[\rho]=1$, imposes
the constraint on the sum of eigenvalues, $\sum_{i=1}^N \ld_i=1$. 

For \emph{random} pure states, the expansion coefficients in eq. \eqref{psi111}
can be typically drawn from an unbiased (so called \emph{Hilbert-Schmidt}) distribution (real or complex)
\begin{equation}\label{HS}
{\rm Prob}[\mathcal{X}]\propto \delta\left( {\rm Tr}(\mathcal{X} \mathcal{X}^{\dagger})-1\right)
\end{equation}
where
the Dyson index $\beta=1,2$ corresponds respectively to real and complex 
$\mathcal{X}$ matrices\footnote{These two cases in turn correspond to quantum systems whose Hamiltonians preserve ($\beta=1$) or break ($\beta=2$) time-reversal symmetry.}. The meaning of eq.
\eqref{HS} is clear: all normalized density matrices compatible with unitary invariance are sampled with equal probability, which corresponds to having minimal \emph{a priori} information about the quantum state under consideration. This in turn induces nontrivial correlations among the Schmidt eigenvalues (which are now real random variables between $0$ and $1$ whose sum is $1$) and makes the investigation of several statistical quantities about such states quite interesting. Here we present a quick summary of known results:
\begin{itemize}
\item the joint probability density (jpd) of Schmidt eigenvalues, derived by Lloyd and Pagels \cite{LP}, which is precisely given by $\mathcal{P}_1(\lambda_1,\ldots,\lambda_N)$
in eq. \eqref{P1}, with $n=N$ and $\alpha=(\beta/2)(N-M+1)-1$. Note that the delta function there guarantees that ${\rm Tr}[\rho]=1$. The normalization constant
in this case reads \cite{ZS}:
\beq
C_{N,\alpha=(\beta/2)(N-M+1)-1}^{(\beta)}=\frac{\Gamma(MN\beta/2)(\Gamma(1+\beta/2))^N}{\prod_{j=0}^{N-1}\Gamma((M-j)\beta/2)\Gamma(1+(N-j)\beta/2)}
\eeq
\item the average von Neumann entropy for large $N,M$ (computed by Page \cite{Page95} for $\beta=2$ and extended in \cite{Arul1} to the case $\beta=1$);
\item the average von Neumann entropy
for {\em finite} $N,M$ and $\beta=2$, conjectured by Page \cite{Page95}
and independently proven by many researchers soon after \cite{proofspage} also in a non-extensive setting \cite{malacarne};
\item density of Schmidt eigenvalues (one-point function) for finite $(N,M)$, derived independently in \cite{densitybeta2} and \cite{adachi} for $\beta=2$
and in \cite{vivodens} for $\beta=1$;
\item universality of eigenvalue correlations for $\beta=2$ \cite{liu};
\item average fidelity between quantum states \cite{zyc} and distribution of so-called $G$-concurrence \cite{cappellini} for $\beta=2$;
\item distribution of so-called {\em purity} for small $N$ \cite{giraud}, and phase transitions in its Laplace transform for large $N$ \cite{scard1};
\item full distribution of R\'enyi entropies (including large deviation tails), computed in \cite{majnadal} for large $N=M$ and all $\beta$s using a Coulomb gas method.
As a byproduct, the authors also obtain in \cite{majnadal} the average and variance of R\'enyi entropy valid for large $N=M$, and the density of Schmidt eigenvalues
for all $\beta$'s and $N=cM$ large;
\item distribution of smallest eigenvalue (related to so-called \emph{Demmel condition number} \cite{demmel}) for $\beta=1,2$ and finite $M=N$, derived independently in \cite{edeldemmel} and \cite{majbohi}. In the latter paper, a conjecture by Znidaric \cite{Znd} was proven
\footnote{For $\beta=2$, $\langle\lambda_{\rm{min}}\rangle=1/N^3$ exactly for all $N=M$, while for $\beta=1$ $\langle\lambda_{\rm{min}}\rangle\sim c_1/N^3$
for large $N=M$, where the constant $c_1$ is precisely known \cite{majbohi}.} (see also \cite{chen} for an extension of these results to the case $N\neq M$);
\item distribution of largest eigenvalue for \emph{large} $N=M$ and all $\beta$s \cite{majnadal}, including small and large deviation laws. Typical fluctuations around the mean $\approx 4/N$
(once properly scaled) are found to follow the Tracy-Widom distribution (see also \cite{nadler} for a related result). No results seem to be available for the case of \emph{finite} $N,M$.
\end{itemize}
Given the current interest in the distribution of extreme Schmidt eigenvalues, the reader may on the one hand wonder whether they really encode useful information, and
on the other why the largest eigenvalue distribution for finite $N,M$ is much harder to obtain via the same strategy used for the smallest one \cite{majbohi}.

In order to answer the first question, first note that due to the constraint $\sum_{i=1}^N \ld_i=1$ and the fact that
all eigenvalues are nonnegative, it follows that\footnote{This means that both the smallest and the largest eigenvalue distributions have compact supports and justifies the bound $x\geq 1/n$ in eq. \eqref{main}.} $1/N\le \ld_{\rm max}\le 1$
and $0\le \ld_{\rm min} \le 1/N$. Now consider the following 
limiting situations. Suppose that the largest eigenvalue $\lambda_{\rm max}=\max_i\{\lambda_i\}$ takes its maximum allowed value $1$. Then 
it follows immediately that all the remaining $(N-1)$ eigenvalues must be identically $0$. In this situation eq. (\ref{Sch1}) tells us that $|\psi\kt$ is fully {\it unentangled} 
(completely separable). On the other hand, if $\lambda_{\rm max}=1/N$ (i.e., it takes its lowest allowed value),
all the eigenvalues must have the same value, $\lambda_i=1/N$ for all $i$. In this case, the pure state $|\psi \kt$ is {\it maximally} entangled, as this state
maximizes the von Neumann entropy $S_{\mathrm{VN}}=-\sum_{i=1}^N\lambda_i\ln\lambda_i=\ln (N)$. In other words, the knowledge of the largest eigenvalue distribution really provides essential information
about how entangled a random pure state is.

A discussion about the asymmetry in the treatment of smallest and largest Schmidt eigenvalues is included in Appendix \ref{appB}.

\subsection{Landauer conductance in open cavities}\label{condcav}
Consider a cavity of submicron dimensions etched in a semiconductor and connected to the external world by two leads supporting $M$
and $N$ electronic channels.
 It is well established that the electrical current flowing
through such a cavity when brought out of equilibrium by an applied external voltage presents time-dependent
fluctuations which persist down to zero temperature
\cite{beenakker} and are thus associated with the granularity of
the electron charge $e$. Typical features
observed in experiments include weak localization
\cite{chang}, universality in conductance fluctuations
\cite{marcus} and constant Fano factor \cite{oberholzer}.
The Landauer-B\"{u}ttiker scattering approach
\cite{beenakker,landauer,buttikerPRL} is rather successful in describing the statistics of quantum transport:
it amounts to relating the wave function
coefficients of the incoming and outgoing electrons
through the unitary scattering matrix $\mathcal{S}$ ($2N_0\times 2N_0$, if $N_0=N+M$):
\begin{equation}\label{ScatteringMatrix S}
  \mathcal{S}=
  \begin{pmatrix}
    \mathbf{r} & \mathbf{t}^\prime \\
    \mathbf{t} & \mathbf{r}^\prime
  \end{pmatrix}
\end{equation}
where the transmission ($\mathbf{t},\mathbf{t}^\prime$) and reflection
$(\mathbf{r},\mathbf{r}^\prime)$ blocks are submatrices encoding the
transmission and reflection coefficients among different channels\footnote{($\mathbf{t},\mathbf{t}^\prime$) are respectively
of size $N\times M$ and $M\times N$, while $(\mathbf{r},\mathbf{r}^\prime)$ are of size $M\times M$ and $N\times N$.}.
Experimental observables can be extracted
from the eigenvalues of the hermitian transport matrix $\mathbf{T}=\mathbf{t} \mathbf{t}^\dagger$: for
example, the dimensionless conductance and the shot noise are
given respectively by $G=\Tr(\mathbf{T})$ \cite{landauer} and
$P=\Tr[\mathbf{T}(\mathbf{1}-\mathbf{T})]$ \cite{ya}.

Random Matrix Theory (RMT) has been very successful in describing the statistics of universal fluctuations in such systems,
when the corresponding classical dynamics is chaotic:
the scattering matrix $\mathcal{S}$ is drawn from a suitable
ensemble of random matrices, with the overall constraint of
unitarity \cite{muttalib,stone,mellopereira}. A
maximum entropy approach (under the assumption of ballistic
point contacts \cite{beenakker}) forces the probability distribution of
$\mathcal{S}$ to be uniform within the unitary group, i.e. $\mathcal{S}$ belongs to
one of Dyson's Circular Ensembles \cite{Mehta,Dys:new}.

From the uniformity of the probability density of $\mathcal{S}$ within the unitary group, the 
jpd of the transmission eigenvalues $\{T_i\}$ of the
matrix $\mathbf{T}$, from which the statistics of interesting experimental quantities could be in principle derived, is precisely given 
by eq. \eqref{P2}, with $n=\min(N,M)$ and $\alpha^\prime = \frac{\beta}{2}(|N-M|+1)-1$ \cite{beenakker,mellopereira,forrcond}.
There, the Dyson index $\beta$ characterizes different symmetry
classes ($\beta=1,2$ according to the presence or absence of
time-reversal symmetry). The eigenvalues $T_i$ are thus correlated real random variables
between $0$ and $1$. The normalization constant $K_{n,\alpha^\prime}^{(\beta)}$ is explicitly known from
the celebrated Selberg's integral as:
\beq
K_{n,\alpha^\prime}^{(\beta)}=\prod_{j=0}^{n-1}\frac{\Gamma(1+\beta/2)\Gamma(2+\alpha^\prime+(\beta/2)(n+j-1))}
{\Gamma(1+(\beta/2)(1+j))\Gamma(1+\alpha^\prime+\beta j/2)\Gamma(1+\beta j/2)}
\eeq

From \eqref{P2}, in principle the
statistics of many observables of interest can be calculated. In particular, we focus here on the dimensionless Landauer conductance $G=\Tr(\mathbf{T})=\sum_{i=1}^n T_i$,
which satisfies the bounds $0\leq G\leq n$.
Mean and variance of $G$ and other quantities have been known for a long time when $N,M$ are large \cite{beenakker,brouwer,novaes}, and recently also for finite $N,M$ \cite{novaes,savin,savin2,vivovivo}. 

Stimulated by some recent experimental progresses \cite{hemmady}, which
made eventually possible to explore the {\em full distribution} of the
conductance (and not just its mean and variance), a lot of effort has been devoted to its theoretical characterization. An
explicit expression was first obtained for $n=1,2$
\cite{baranger,jalabert,garcia}, while more results were available
in the case of quasi one-dimensional wires \cite{muttalibwolfle}
and 3D insulators \cite{muttalib3D}. For the shot noise, the
full distribution was known only for $N=M=1$ \cite{pedersen}. Very
recently, Sommers {\it et al.} \cite{sommers} announced two
formulas for the distribution of conductance and
shot noise, valid at arbitrary number of open channels and for any $\beta$, which are
based on Fourier expansions. In \cite{savinnew}, a complete and systematic approach to the full statistics of conductance and shot noise was brought forward. This is based
on symmetric function expansions, and is valid for $\beta=1,2$ and arbitrary $M,N$. In addition, the authors of \cite{savinnew} were able to provide
general formulae in terms of determinants or Pfaffians for the full probability distribution of conductance and shot noise at quantized $\beta$ and general $N,M$ (see below).
In \cite{Kanz}, the integrable theory of quantum transport in chaotic cavities (based on Painlev\'e transcendent) for $\beta=2$ was formulated, and recursion formulae for the efficient computation of conductance and shot noise cumulants have been derived. In two recent publications \cite{vivoPRL}, the distributions of conductance, shot noise and integer moments were computed in the limit $N=M\gg 1$
using a Coulomb gas method, and long {\em power-law} tails were detected
in the distributions\footnote{A careful asymptotic analysis of formulae in \cite{Kanz} leads to the same findings. In particular, there are no long exponential tails in the distributions, as originally claimed in \cite{Kanz}.}, a result confirmed
by extensive numerical simulations.
In \cite{kumar}, exact results (basically equivalent to the general formulae in \cite{savinnew}) for the Laplace transform $\tilde{\mathcal{P}}_G(s)$ of the conductance distribution for any $N,M$ and all $\beta$s are formally given as follows:
\begin{itemize}
\item \underline{$\beta=1$}: $\tilde{\mathcal{P}}_G(s)\propto\mathrm{Pf}[\Psi^{(1)}_{j,k}(s)]_{j,k=0,\ldots,N-1}$ for $N$ even, and
$\tilde{\mathcal{P}}_G(s)\propto\mathrm{Pf}\left[\begin{array}{cr}
			\Psi^{(1)}_{j,k}(s) & \Phi^{(1)}_{j}(s)     \\
			-\Phi^{(1)}_{k}(s)    & 0  
		\end{array}\right]_{j,k=0,\ldots,N-1}$ for $N$ odd, where $\mathrm{Pf}[\mathcal{A}]$ is the Pfaffian of the even-dimensional antisymmetric matrix
		$\mathcal{A}$ \cite{Mehta} and the proportionality constants are known explicitly. The arrays $\Psi^{(1)}_{j,k}(s)$ and $\Phi^{(1)}_{j}(s)$ are
		given by:
		\begin{align}
		\Psi^{(1)}_{j,k}(s) &=\int_0^1\int_0^1\mathrm{sgn}(x-y)e^{-sx}e^{-sy}x^{\alpha^\prime +j}y^{\alpha^\prime +k}dx\ dy\\
		\Phi^{(1)}_{j}(s)  &=\int_0^1 e^{-sx}x^{\alpha^\prime +j}dx
		\end{align} 
\item \underline{$\beta=2$}: in this case, we have a representation in terms of a Hankel determinant 
\beq\tilde{\mathcal{P}}_G(s)\propto\det[\Psi^{(2)}_{j,k}(s)]_{j,k=0,\ldots,N-1}\label{hankel}\eeq
where:
\begin{equation}
\Psi^{(2)}_{j,k}(s)=\int_0^1 e^{-s x}x^{\alpha^\prime +j+k}dx
\end{equation}
that was first derived in \cite{Kanz}.
\end{itemize}
Explicit inversions of the Laplace transforms above are always possible on a case-by-case basis, and a catalogue of such evaluations for few interesting cases is
provided in ref. \cite{kumar}. In next section and in Appendix \ref{appC}, we will combine such explicit formulas with the identity in eq. \eqref{main} to illustrate the validity of our approach
to the distribution of largest Schmidt eigenvalue.

\section{Two applications of the main identity}\label{mainresults}
For illustrative purposes, we consider two applications of the main identity eq. \eqref{main}. A third one is discussed in great detail in appendix \ref{appC}.
\begin{enumerate}
\item \underline{$N=3,M=4,\beta=1$}:
in this case, the probability density of Landauer conductance has been derived explicitly in \cite{kumar} as:
\begin{align}
\nonumber \mathcal{P}_G(y) &=\frac{3}{8}\left[y^5-(y-1)^3 (y^2-12 y+51)\theta(y-1)-(y-2)^3\right.\\
&\left.\times (y^2+6 y+24)
\theta(y-2)\right]\theta(3-y)\label{pptt}
\end{align}
where $\theta(x)$ is Heaviside step function. 
\item\underline{$N=3,M=5,\beta=1$}:
again, the probability density of Landauer conductance in this case has been derived explicitly in \cite{kumar} as:
\begin{equation}
\mathcal{P}_G(y) =
\begin{cases}
\frac{20}{143}y^{13/2} & \mbox{for }0\leq y\leq 1\\
\frac{5}{2288}\left[3003y^5-21021y^4+55770y^3\right. & \\
-70070y^2+42315y-9933 &\\
 -32(y-2)^{7/2}(2 y^3+14 y^2+63 y+231) &\\
\left.\times\theta(y-2)\right] &\mbox{for }1\leq y\leq 3
\end{cases}\label{pptt1}
\end{equation}
\end{enumerate}

Note that:
\begin{itemize}
\item since the conductance density has always a compact support $[0,n]$ (where $n=\min(N,M)$), it follows immediately
from eq. \eqref{main} that the cumulative distribution of the largest Schmidt eigenvalue $Q_n(x)$ and its density $p_n(x)$ have compact support $1/n\leq x\leq 1$,
as expected.
\item since the conductance density is known to be continuous but not everywhere analytic \cite{sommers,savinnew} (i.e. it displays 'critical' points at which higher derivatives are discontinuous), the cumulative distribution and the density
of the largest Schmidt eigenvalue enjoy this property too.
\end{itemize}

\begin{figure}
\begin{center}
\includegraphics[bb=0 0 240 158,width=.7\hsize]{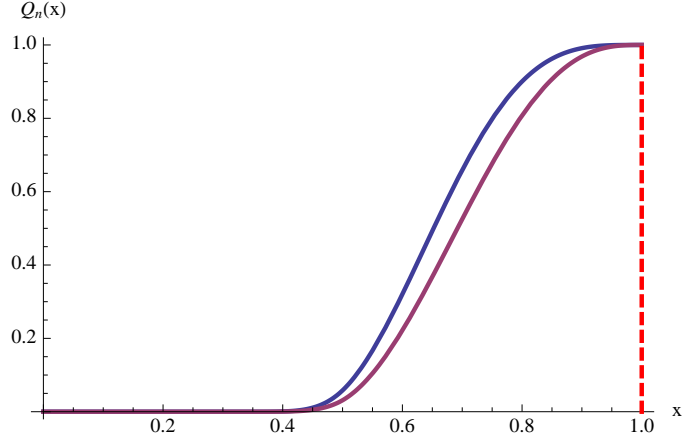}
\caption{Cumulative distribution $Q_n(x)$ of the largest Schmidt eigenvalue via eq. \eqref{main} for $n=N=3,M=4,\beta=1$ (violet) and $n=N=3,M=5,\beta=1$ (blue). This distribution is identically zero for $x\leq 1/n$ and $Q_n(1)=1$ as it should.} \label{cumfig}
\end{center}
\end{figure}

\begin{figure}
\begin{center}
\includegraphics[bb= -269   112   882   680,width=.9\hsize]{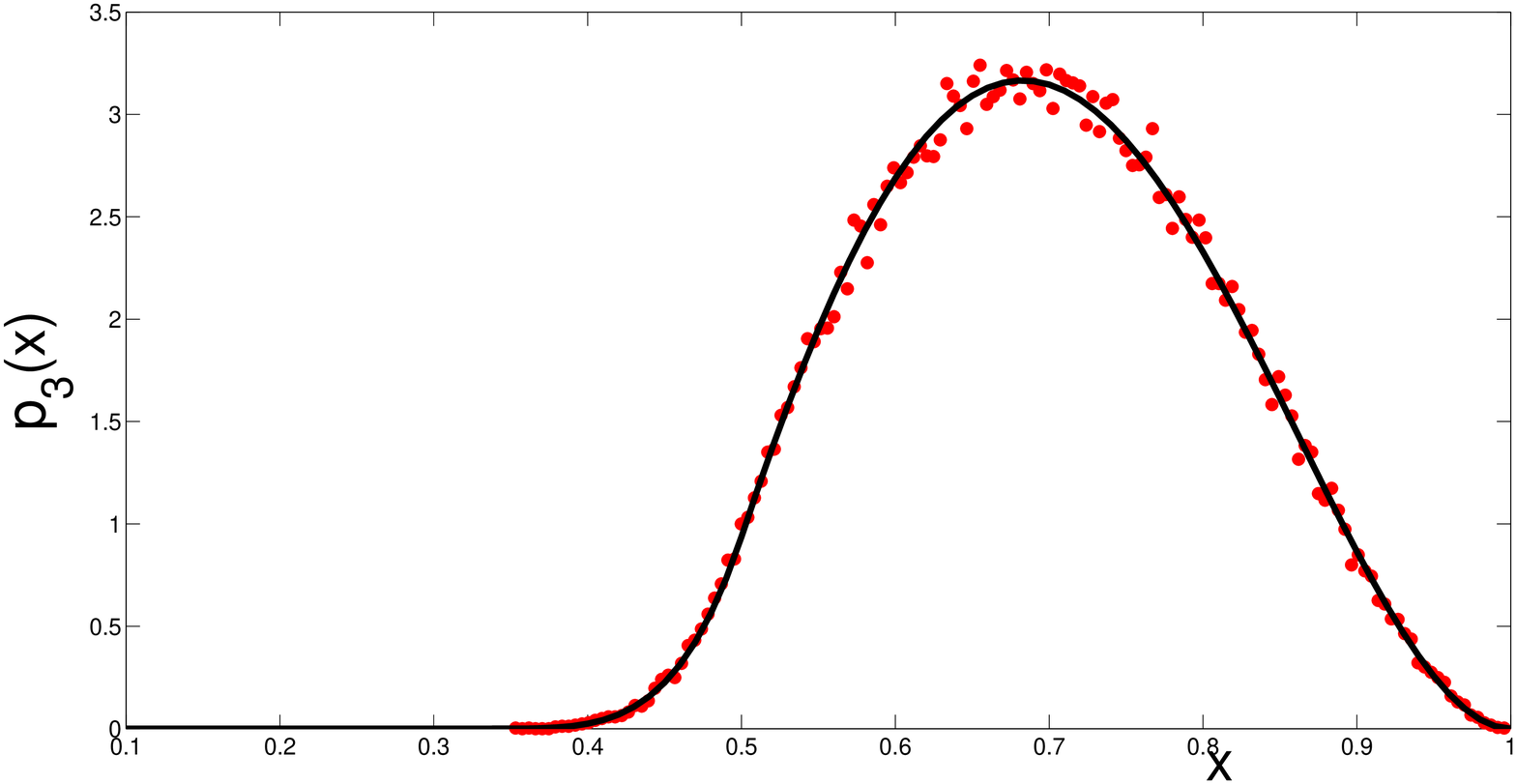}
\caption{Density of the largest Schmidt eigenvalue $p_n(x)$ via eqs. \eqref{densq} and \eqref{main} for $n=N=3,M=4,\beta=1$ (solid black curve), along with
numerical diagonalization of $\kappa=8\cdot 10^4$ samples (red dots), see main text for the algorithm.} \label{densityfig1}
\end{center}
\end{figure}

\begin{figure}
\begin{center}
\includegraphics[bb= -269   112   882   680,width=.9\hsize]{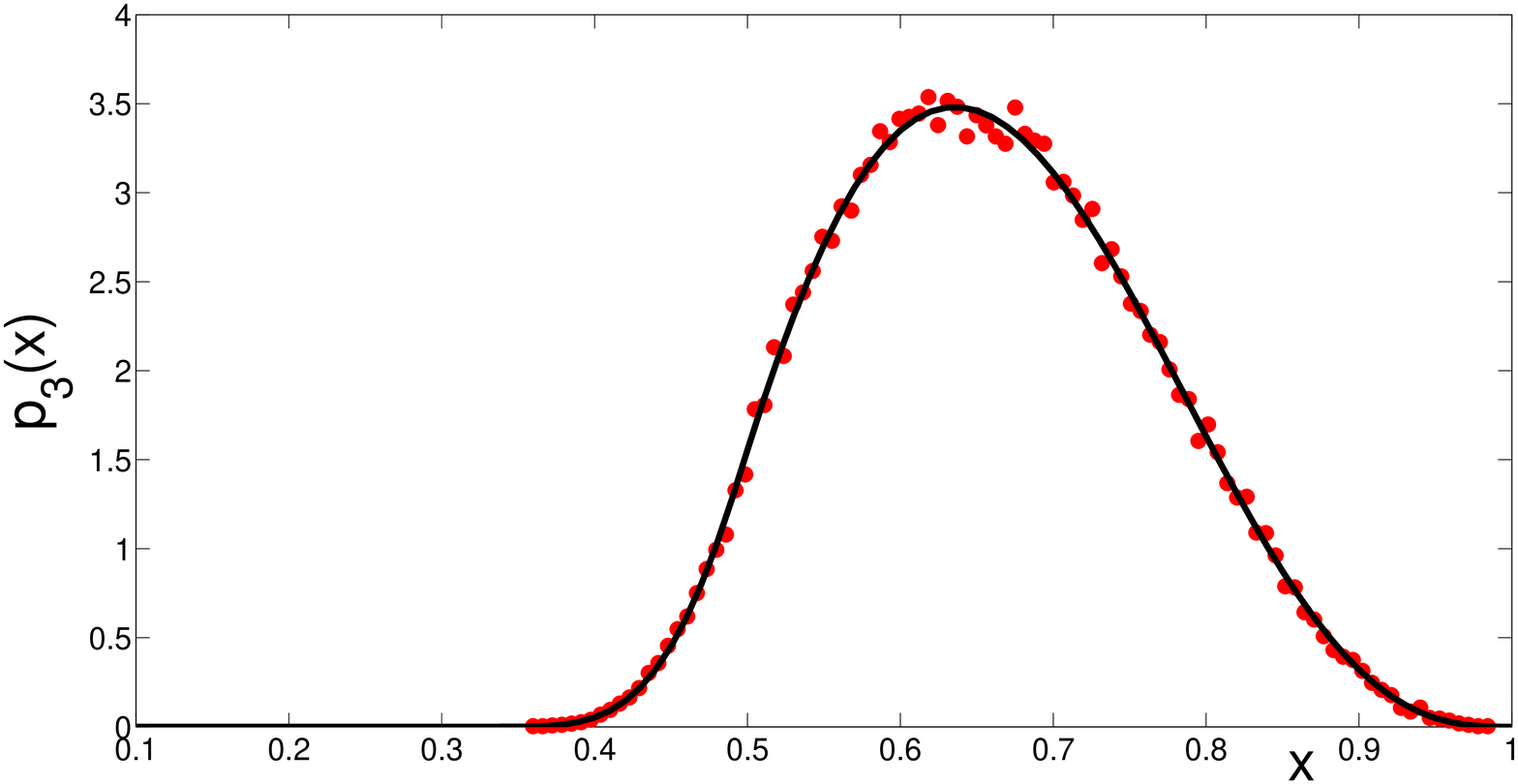}
\caption{
Density of the largest Schmidt eigenvalue $p_n(x)$ via eqs. \eqref{densq} and \eqref{main} for $n=N=3,M=5,\beta=1$ (solid black curve), along with
numerical diagonalization of $\kappa=10^5$ samples (red dots), see main text for the algorithm.} \label{densityfig2}
\end{center}
\end{figure}

In fig. \ref{cumfig} we plot the cumulative distributions $Q_n(x)$ corresponding to the two cases above. The curves are obtained via eq. \eqref{main}, where $\mathcal{P}_G(y)$ is respectively given
by \eqref{pptt} and \eqref{pptt1}. The distributions are increasing functions of the argument $x$, as they should, and such that $Q_n(1)=1$. Differentiation of the analytical formulas
provide the density of the largest eigenvalue, $p_n(x)$, which is plotted in figs. \ref{densityfig1} and \ref{densityfig2} along with numerical simulations.
These are obtained as follows \cite{ZS,ZyckBook}:
\begin{enumerate}
\item we generate $\kappa\simeq 10^4,10^5$ {\em real} Gaussian $M\times N$ matrices $\mathcal{X}$.
\item for each instance we construct the
Wishart matrix $\mathcal{W}=\mathcal{X}^T \mathcal{X}$. 
\item we diagonalize $\mathcal{W}$ and collect its $N$ real and non-negative eigenvalues $\{ \tilde{\lambda}_1,\ldots,\tilde{\lambda}_N\}$.
\item we define a new set of variables $0\leq \lambda_i\leq 1$ as $\lambda_i =\tilde{\lambda}_i /\sum_{i=1}^N \tilde{\lambda}_i$, for $i=1,\ldots,N$.
The set of variables $\lambda_i$ is guaranteed to be sampled according to the measure (\ref{P1}).
\item we construct a normalized histogram of $\lambda_{\rm{max}}=\max_i\{\lambda_i\}$.
\end{enumerate}
The agreement between theory and simulations is excellent. 

We also evaluated exactly the average of $\langle\lambda_{\rm max}\rangle=\int_{1/n}^1 dx\ x\ p_n(x)$ for the two cases above and found:
\begin{align}
\langle\lambda_{\rm max}\rangle_1 &=\frac{25}{36} \approx 0.694444...\\
\langle\lambda_{\rm max}\rangle_2 &=\frac{1}{810}(378+89\sqrt{3}) \approx 0.656978...
\end{align}
However, a general formula for $\langle\lambda_{\rm max}\rangle$ is still elusive (unlike $\langle\lambda_{\rm min}\rangle$ \cite{majbohi}) and may well attract further researches (see also Appendix \ref{appC}).

\section{Conclusions}\label{concl}
We have presented an exact identity relating two statistical quantities which arise in different contexts: the cumulative distribution of the largest Schmidt eigenvalue
for entangled random pure states in bipartite systems of sizes $(M,N)$ and the probability density of Landauer conductance in chaotic cavities supporting
$N$ and $M$ electronic channels in the two external leads. Recent analytical results for the latter are exploited to derive (so far unavailable) exact formulas for
the former quantity at {\em finite} $N,M$ (while large $N,M$ results are already available \cite{majnadal}), which is of interest in order to quantify the degree of entanglement of random pure states. A detailed introduction to the physics involved
has been provided, along with a precise discussion of the asymmetry in the treatment of the smallest and largest Schmidt eigenvalue distributions. A general formula
for $\langle\lambda_{\rm{max}}\rangle$, the average of the largest eigenvalue, valid for arbitrary $N,M$ is unfortunately still lacking and certainly deserves further investigations.\\
\\
{\bf Acknowledgments:} it is a pleasure to thank Satya N. Majumdar, C\'eline Nadal, Oriol Bohigas and Gernot Akemann for collaborations on related projects and many interesting discussions, and Boaz Nadler for useful correspondence. I am also grateful to Dima Savin for a careful reading of the manuscript and many helpful advices.

\appendix
\section{Derivation of main identity.}\label{appA}

Consider the cumulative distribution $Q_n(x)$ of $\lambda_{\mathrm{max}}=\max_i\{\lambda_i\}$ (eq. \eqref{cum1}):
\begin{equation}
Q_n(x)=C_{n,\alpha}^{(\beta)}\int_{[0,x]^n}d\lambda_1\cdots d\lambda_n \delta\left(\sum_{i=1}^n\lambda_i -1\right)\prod_{i=1}^n
\lambda_i^{\alpha}\prod_{j<k}|\lambda_j-\lambda_k|^\beta
\end{equation}
A change of variables $\lambda_i=x T_i$ leads to:
\begin{align}
\nonumber Q_n(x) &=C_{n,\alpha}^{(\beta)} x^{n+\alpha n+\frac{\beta}{2}n (n-1)}\times\\
&\times\int_{[0,1]^n}d T_1\cdots d T_n 
\delta\left(x\sum_{i=1}^n T_i -1\right)\prod_{i=1}^n T_i^{\alpha}\prod_{j<k}|T_j-T_k|^\beta
\end{align}
Using the property of delta functions $\delta(\gamma Y)=\delta(Y)/|\gamma|$, we obtain straighforwardly eq. \eqref{main}.

\section{Asymmetry in the treatment of smallest and largest Schmidt eigenvalues}\label{appB}

Consider the cumulative distribution $\Theta_n(x)$ of the \emph{smallest} member of set $1$, $\lambda_{\rm min}=\min_i \{\lambda_i\}$ (i.e. the distribution
of the smallest Schmidt eigenvalue for entangled random pure states). By definition it is given by:
\begin{equation}\label{cummin}
\Theta_n(x)=C_{n,\alpha}^{(\beta)}\int_{[x,\infty]^n}d\lambda_1\cdots d\lambda_n \delta\left(\sum_{i=1}^n\lambda_i -1\right)\prod_{i=1}^n
\lambda_i^{\alpha}\prod_{j<k}|\lambda_j-\lambda_k|^\beta
\end{equation}
where the upper limit of integration can be safely extended up to $\infty$ in view of the unit norm constraint.
In the case $\alpha=0$ and $\beta=2$, which we focus on here for illustrative purposes, the evaluation of this multiple integral proceeds via the auxiliary function $\Theta_n(x,t)$ \cite{majbohi}:
\begin{equation}\label{cummint}
\Theta_n(x,t)=C_{n,0}^{(2)}\int_{[x,\infty]^n}d\lambda_1\cdots d\lambda_n \delta\left(\sum_{i=1}^n\lambda_i -t\right)
\prod_{j<k}|\lambda_j-\lambda_k|^2
\end{equation}
such that $\Theta_n(x)\equiv \Theta_n(x,1)$. Next, one takes the Laplace transform of $\Theta_n(x,t)$:
\begin{equation}\label{cummint}
\int_0^\infty dt\Theta_n(x,t)e^{-st}=C_{n,0}^{(2)}\int_{[x,\infty]^n}d\lambda_1\cdots d\lambda_n e^{-s\sum_{i=1}^n\lambda_i }
\prod_{j<k}|\lambda_j-\lambda_k|^2
\end{equation}
and in the r.h.s performs a linear shift $y_i=s(\lambda_i-x)$ (this is the crucial technical step), to get:
\begin{equation}\label{cummint}
\int_0^\infty dt\Theta_n(x,t)e^{-st}=\frac{e^{-sNx}}{s^{N^2}}C_{n,0}^{(2)}\underbrace{\int_{[0,\infty]^n}d y_1\cdots d y_n e^{-\sum_{i=1}^n y_i }
\prod_{j<k}|y_j-y_k|^2}_{\propto\mathcal{Z}_{\rm{WL}}}
\end{equation}
Note that, thanks to the linear shift, the dependence on the Laplace variable $s$ has been entirely transferred outside the $n$-fold integral: this is
now proportional to the partition function $\mathcal{Z}_{\rm{WL}}$ of an associated Wishart-Laguerre (WL) ensemble of random covariance matrices
\cite{Mehta,Wishart} of the form $\mathcal{W}=\mathcal{X}^\dagger\mathcal{X}$, where $\mathcal{X}$ is a Gaussian rectangular matrix with real or complex entries.
The joint distribution of the $n$ nonnegative eigenvalues of $\mathcal{W}$ is known~\cite{James}
\beq
\mathcal{P}^{(\rm{WL})}(\lambda_1,\ldots,\lambda_n) =\mathcal{N}_{n,\alpha}^{(\beta)}\, e^{-\frac{\beta}{2}\sum_{i=1}^n 
\lambda_i}\, \prod_{i=1}^n \lambda_i^{\alpha}\, \prod_{j<k} 
|\lambda_j-\lambda_k|^{\beta}
\label{wishart1}
\eeq
where $\mathcal{N}_{n,\alpha}^{(\beta)}$ is a known normalization constant. Therefore, the jpd of Schmidt eigenvalues \eqref{P1} can be seen as a {\em fixed-trace} (microcanonical) version of the Wishart-Laguerre (canonical) ensemble\footnote{Note that the presence of a fixed-trace
constraint has crucial consequences on the spectral properties of random matrix ensembles \cite{akemann,ASOS}.}. 

Now, in the case of the cumulative distribution of the \emph{largest} eigenvalue $Q_n(x)$, the integrals on the r.h.s. run over $[0,x]$ instead of $[x,\infty]$, making 
the aforementioned linear shift less useful. One could keep pursuing the Laplace-transform route (introducing an auxiliary function $Q_n(x,t)$) with the change of variables $s\lambda_i=y_i$ obtaining:
\begin{equation}
\int_0^\infty dt Q_n(x,t)e^{-st}=\frac{C_{n,0}^{(2)}}{ s^{n^2}}\underbrace{\int_{[0,sx]^n}d y_1\cdots d y_n e^{-\sum_{i=1}^n y_i }
\prod_{j<k}|y_j-y_k|^2}_{\propto E_{\rm{WL}}[(sx,\infty)]}
\end{equation}
but the integral on the r.h.s. does not permit this time a friendly Laplace-inversion (see however next appendix). Indeed, this integral is readily recognized as proportional to $E_{\rm{WL}}[(sx,\infty)]$,
where $E_{\rm{WL}}[(a,b)]$ is the \emph{gap probability} for a Wishart-Laguerre ensemble, i.e. the probability that the interval $(a,b)$ on the real axis is free of eigenvalues.
This quantity is exactly known in terms of Painlev\'e V \cite{TW}, with the consequence that an explicit Laplace inversion formula is not available to date.

In summary, the asymmetry in the treatment of the smallest and largest Schmidt eigenvalues arises because the linear shift that works in the former case fails in the latter, and this fact calls for the alternative approach developed in this paper.

\section{A third application of main identity for $\beta=2$ and $N=M$}\label{appC}

In this appendix we discuss in more detail a third application of the main identity eq. \eqref{main}. In the case $\beta=2$ and $M=N$,
the Hankel determinant representation in eq. \eqref{hankel} actually allows a more systematic (and more easily automatized) treatment of the cumulative distribution of the largest Schmidt eigenvalue.
Following the Laplace transform route outlined in the previous appendix, one can easily write down the following equation:
\beq
\int_0^\infty dt Q_n(x,t)e^{-st/x}=C_{n,0}^{(2)}\ x^{n^2}\underbrace{\int_{[0,1]^n}d y_1\cdots d y_n e^{-s\sum_{i=1}^n y_i}\prod_{j<k}|y_j-y_k|^2}_{n!\ \det[(-\partial_s)^{j+k}\mathcal{F}_1(s)]_{j,k=0,\ldots, n-1}}
\eeq
where we have used the Hankel determinant representation for the $n$-fold integral on the r.h.s. provided in \cite{Kanz} (fully equivalent to \eqref{hankel}), with $\mathcal{F}_1(s)=(1-e^{-s})/s$.

\begin{figure}
\begin{center}
\includegraphics[bb=0 0 240 160,width=.7\hsize]{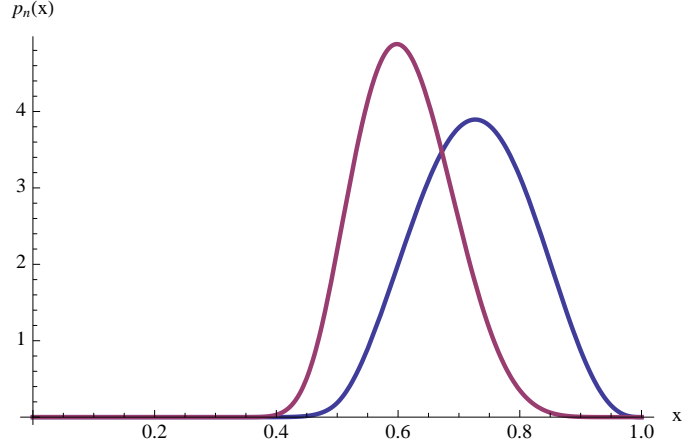}
\caption{Probability density $p_n(x)=\frac{d}{dx}Q_n(x)$ of the largest Schmidt eigenvalue via eq. \eqref{invlap} for $n=N=M=3$ (blue) and $n=N=M=4$ (violet) and $\beta=2$. } \label{densityapp}
\end{center}
\end{figure}

A change of variable $t=x\tau$ on the l.h.s. followed by a (formal) Laplace inversion, leads to the final formula for $Q_n(x)$ in this case\footnote{One could have started directly from eq. \eqref{main}
and would have been led to the very same eq. \eqref{invlap} upon noticing that $\mathcal{P}_G(1/x)=(n!/K_{n,0}^{(2)})\mathcal{L}^{-1}\left[\det[(-\partial_s)^{j+k}\mathcal{F}_1(s)]_{j,k=0,\ldots, n-1}\right](1/x)$ \cite{Kanz}.}:
\beq
Q_n(x)=C_{n,0}^{(2)}\ n!\ x^{n^2-1}\mathcal{L}^{-1}\left[\det[(-\partial_s)^{j+k}\mathcal{F}_1(s)]_{j,k=0,\ldots, n-1}\right](1/x)\label{invlap}
\eeq
where $\mathcal{L}^{-1}[f(s)](t)$ is the inverse Laplace transform of $f(s)$ with parameter $t$. The r.h.s. of \eqref{invlap} can be systematically
evaluated in \textsf{Mathematica}$^{\textregistered}$ and its derivative ($p_n(x)$) has been plotted in fig. \ref{densityapp} for $n=N=M=3,4$.

Also the average of the largest eigenvalue $\langle\lambda_{\rm{max}}\rangle=\int_{1/n}^1 dx\ x\ p_n(x)$ can be evaluated exactly in \textsf{Mathematica}$^{\textregistered}$
for a given $n$, and we provide a few evaluations in the following table. Unlike the average of the smallest eigenvalue, which in the same circumstances ($\beta=2$ and $M=N$) has the attractively simple form 
$\langle\lambda_{\rm{min}}\rangle = 1/N^3$ exactly for all $N$ \cite{majbohi}, the situation for the largest eigenvalue seems more complicated and it is much harder to even conjecture
a possible formula for the sought average valid for all $N=M$. We leave this as a challenging open problem, noticing {\em en passant} that the asymptotic value $\langle\lambda_{\rm{max}}\rangle \sim 4/N$
derived in \cite{majnadal} is approached {\em very slowly} as $N$ increases \cite{vivodens}. This fact makes the investigation of {\em finite} (small) $N$ results very much called for.

\begin{center}
\begin{tabular}{ | c | c| c| }
  \hline                       
  $N=M$ & $\langle\lambda_{\rm{max}}\rangle$ (exact) & $\langle\lambda_{\rm{max}}\rangle$ (approx.)\\ \hline
  & & \\
  $2$ & $\frac{7}{8}$ & $0.875$ \\
  & & \\
  $3$ & $\frac{313}{432}$ & $0.724537$\\ 
  & & \\
 $4$ & $\frac{1 367 807}{2 239 488}$ & $0.610768$\\ 
 & & \\
  $5$ & $\frac{4 581 882 694 877}{8 707 129 344 000}$ & $0.526222$\\ 
  & & \\ 
  $6$ & $\frac{225 128 892 964 655 720 357 665 283}{487 487 792 008 396 800 000 000 000}$ & $0.461814$\\
  & &\\ \hline
\end{tabular}
\end{center}

\end{document}